\documentclass{aa}  
\usepackage[varg]{txfonts}
\usepackage{natbib}
\usepackage{graphicx}
\newcommand{\rev}{}

\begin{document}
\title{$\gamma$ Peg: testing Vega-like magnetic fields in B
stars\thanks{Based on observations obtained at the Telescope Bernard Lyot
(USR5026) operated by the Observatoire Midi-Pyr\'en\'ees, Universit\'e de
Toulouse (Paul Sabatier), Centre National de la Recherche Scientifique of
France, and at the Dominion Astrophysical Observatory.}}
\titlerunning{$\gamma$ Peg: testing Vega-like fields}

\author{C. Neiner\inst{1}
\and D. Monin\inst{2}
\and B. Leroy\inst{1}
\and S. Mathis\inst{3,1}
\and D. Bohlender\inst{2}
}

\offprints{C. Neiner}

\institute{LESIA, Observatoire de Paris, CNRS UMR 8109, UPMC, Universit\'e Paris Diderot, 5 place Jules Janssen, 92190 Meudon, France; 
\email{coralie.neiner@obspm.fr}
\and Dominion Astrophysical Observatory, Herzberg Astronomy and Astrophysics Program, National Research Council of Canada, 5071 West Saanich Road, Victoria, BC V9E 2E7, Canada
\and Laboratoire AIM Paris-Saclay, CEA/DSM-CNRS-Universit\'e Paris Diderot;
IRFU/SAp, Centre de Saclay, 91191 Gif-sur-Yvette Cedex, France
}

\date{Received 20 November 2013; accepted 11 December 2013}
 
\abstract
{$\gamma$\,Peg is a bright B pulsator showing both p and g modes of $\beta$\,Cep
and SPB types. It has also been claimed to be a magnetic star by some authors
while others do not detect a magnetic field.}
{We aimed at checking for the presence of a magnetic field, characterise it if
it exists or provide a firm upper limit of its strength if it is not detected.
If $\gamma$\,Peg is magnetic as claimed, it would make an ideal
asteroseismic target to test various theoretical scenarios. If it is very weakly
magnetic, it would be the first observation of an extension of Vega-like fields
to early B stars. Finally, if it is not magnetic and we can provide a very low
upper limit on its non-detected field, it would make an important result for
stellar evolution models.}
{We acquired  high resolution, high signal-to-noise spectropolarimetric Narval
data at Telescope Bernard Lyot (TBL). We also gathered existing dimaPol
spectropolarimetric data from the Dominion Astrophysical Observatory (DAO) and
Musicos spectropolarimetric data from TBL. We analysed the Narval and Musicos
observations using the LSD (Least-Squares Deconvolution) technique to derive the
longitudinal magnetic field and Zeeman signatures in lines. The longitudinal
field strength was also extracted from the H$\beta$ line observed with the DAO.
With a Monte Carlo simulation we derived the maximum strength of the field
possibly hosted by $\gamma$\,Peg.}
{We find that no magnetic signatures are visible in the very high quality
spectropolarimetric data. The average longitudinal field measured in the Narval
data is $B_l=-0.1\pm0.4$ G. We derive a very strict upper limit of the dipolar
field strength of $B_{\rm pol}\sim40$ G.}
{We conclude that $\gamma$\,Peg is not magnetic: it does not host a strong
stable fossil field as observed in a fraction of massive stars, nor a very weak
Vega-like field. There is therefore no evidence that Vega-like fields exist in B
stars contrary to the predictions by fossil field dichotomy scenarios. These
scenarios should thus be revised. Our results also provide strong constraints
for stellar evolution models.}


\maketitle

\section{Introduction}\label{intro}

$\gamma$\,Peg is a very bright (V=2.83) B2IV star, which hosts both p and g
pulsation modes. From space-based MOST observations, \cite{handler2009} detected
eight $\beta$\,Cep-like p-modes and six SPB-like g-modes, with frequencies
ranging from 0.6 to 9.1 c~d$^{-1}$. \cite{Walczak2010} provide the most probable
identification for these modes. In addition, $\gamma$\,Peg is an intrinsically
very slow rotator, with vsini close to 0 km~s$^{-1}$ \citep{telting2006} and
v$\sim$3 km~s$^{-1}$ \citep{handler2009}. Moreover, $\gamma$\,Peg has been
claimed to be a {\rev multiple star \citep{chapellier2006}, but
\cite{mcalister1989} and \cite{roberts2007} did not detect a companion.}
\cite{handler2009} proposed that the observed variations are rather due to the
pulsations and that $\gamma$\,Peg is a single star.

Finally, \cite{butkovskaya2007} (hereafter BP07) claimed the detection of a
magnetic field in $\gamma$\,Peg with a longitudinal field varying from -10 to 30
G with a period $P=6.6538\pm0.0016$ d. This longitudinal field would correspond
to an oblique dipole field with a polar field strength of $B_{\rm pol}$=570 G,
an inclination angle $i$=9$^\circ$ and an obliquity angle $\beta$=85$^\circ$.
\cite{silvester2009}, however, detected no magnetic signature in an ESPaDOnS
measurement and derived a null longitudinal field value of $B_l=6\pm11$ G.
\cite{schnerr2008} also found no magnetic signature in 2 Musicos measurements
with $B_l=3\pm20$ G and $B_l=-1\pm17$ G. In addition, \cite{morel2008} studied
the chemical abundances of magnetic pulsating B stars, particularly their
N-enrichment, and found $\gamma$\,Peg to stand out as a normal N star whereas
other magnetic pulsating B stars seem to have N-enrichment. Moreover,
$\gamma$\,Peg appears as non-variable in the UV wind resonance lines, contrary
to most magnetic massive stars \citep{schnerr2008}. Therefore doubts can be cast
on the detection of a magnetic field in $\gamma$\,Peg by BP07. 

Whether $\gamma$\,Peg hosts a magnetic field is important for several reasons:

First of all, \cite{auriere2007} proposed that dipolar magnetic fields can only
exist above a certain polar field strength threshold. They obtained a plateau at
$\sim$1000 G and a threshold at $\sim$300 G. This critical field value would be
necessary for the stability of large-scale magnetic fields. If the field
suggested by BP07 exists, it would fall in this category. The non-detection of a
magnetic field in $\gamma$\,Peg by other authors, however, suggests that if
$\gamma$\,Peg is magnetic its field might be very weak. \cite{lignieres2009}
discovered a field well below the critical field limit with $B_l=-0.6\pm0.3$ G
in the A star Vega \citep[also see][]{petit2010} and \cite{petit2011} discovered
a field with $B_l=0.2\pm0.1$ G in the Am star Sirius. These authors suggest that
a new class of very weakly magnetic stars may exist among intermediate-mass and
massive stars. A dichotomy would then exist between these very weakly magnetic
stars, for which the longitudinal field is in the sub-Gauss regime, and the
dipolar magnetic stars with field strength above $\sim$300 G at the poles. It is
thus interesting to check whether $\gamma$\,Peg could be a B counterpart of this
new "Vega-like" category. 

Second, the presence of a magnetic field in massive stars modifies 
significantly their evolution. In particular the interaction between rotation
and a magnetic field may completely modify the transport of angular momentum and
of chemical elements. If the field suggested by BP07 exists, it would inhibit
mixing in $\gamma$\,Peg
\citep[e.g.][]{spruit1999,mathis2005,zahn2011,briquet2012}. If $\gamma$\,Peg
hosts a very weak field, its impact on transport will probably depend on the
nature of the field, its {\rev strength} and complexity inside the star. The
effect of such weak fields would have to be investigated theoretically
\citep[see the discussion in][]{zahn2007}. Therefore, it is necessary to know
whether very weak fields are present in upper main sequence stars and should be
included in stellar evolution codes \citep[e.g.][]{maeder2003,heger2005}. 

In this paper, we investigate new Narval measurements of the magnetic field in
$\gamma$\,Peg as well as unpublished archival Musicos and DAO data
(Sect.~\ref{obs}). We measure the longitudinal field values with the H$\beta$
line in the DAO data and with the Least-Square Deconvolution (LSD) technique in
the Musicos and Narval data (Sect.~\ref{analysis}). We then derive the upper
limit of the non-detected field (Sect.~\ref{limit}) and discuss the impact on
this non-detection at a very low field level on fossil field theories and
evolution  models (Sect.~\ref{discuss}).

\section{Observations}\label{obs}

\subsection{Musicos observations}

Musicos is a fibre-fed echelle spectropolarimeter with a resolving power of
35000, which was attached to the 2-meter T\'elescope Bernard Lyot (TBL) at the
Pic du Midi in France until 2006 \citep[see][]{donati1999}. The spectrograph
covers the wavelength domain from 4490 to 6619 \AA. 

We collected 34 Stokes V (circular polarisation) measurements of $\gamma$\,Peg
with Musicos between 2001 and 2005. Their signal-to-noise (S/N) varies between 408 
and 1202 at 5000 \AA\ in the I spectrum. See Table~\ref{logmusicos}.

One Stokes Q+U observation was also obtained on October 9, 2003. However, we
consider here only measurements of circular polarisation. Indeed the signatures
detected in linear polarisation due to a stellar magnetic field in massive stars
is usually 10 to 100 times weaker than the one observed in circular
polarisation.

\begin{table}[!ht]
\caption[]{Journal of 34 Musicos observations of $\gamma$\,Peg. The S/N given in
col. 5 is the one measured at 5000 \AA\ in the I spectrum. }
\begin{center}
\begin{tabular}{llllrrr}
\hline
\hline
\# & Date & mid-HJD	& T$_{\rm exp}$ & S/N & $B_l$  & $\sigma_{B_l}$  \\  
   & 	  & $-$2450000  & s		&     & G      & G  \\
\hline
1   & 19dec01 & {\rev 2263.27782} & 4$\times$300 & {\rev  950} &   9.4 &   6.9 \\
2   & 19dec01 & {\rev 2263.29387} & 4$\times$300 & {\rev  840} &   4.0 &   7.6 \\
3   & 20jun02 & {\rev 2446.65110} & 4$\times$180 & {\rev  460} &  12.9 &  11.9 \\
4   & 27jun02 & {\rev 2453.64276} & 4$\times$180 & {\rev  580} &  -2.9 &  10.0 \\
5   & 15nov04 & {\rev 3325.41329} & 4$\times$360 & {\rev  730} &  -6.0 &   7.9 \\
6   & 16nov04 & {\rev 3326.38272} & 4$\times$600 & {\rev  860} &  -4.2 &   7.1 \\
7   & 17nov04 & {\rev 3327.40198} & 4$\times$600 & {\rev  880} &  -6.3 &   6.9 \\
8   & 18nov04 & {\rev 3328.42364} & 4$\times$600 & {\rev  990} &  -0.7 &   6.2 \\
9   & 20nov04 & {\rev 3330.35798} & 4$\times$600 & {\rev  920} &   5.5 &   6.6 \\
10  & 21nov04 & {\rev 3331.39483} & 4$\times$600 & {\rev 1030} &  -2.9 &   6.1 \\
11  & 22nov04 & {\rev 3332.27740} & 4$\times$600 & {\rev  410} &  -1.0 &  13.4 \\
12  & 23nov04 & {\rev 3333.38135} & 4$\times$600 & {\rev 1060} &  -1.0 &   6.0 \\
13  & 24nov04 & {\rev 3334.32173} & 4$\times$600 & {\rev  410} & -38.2 &  47.2 \\
14  & 25nov04 & {\rev 3335.37968} & 4$\times$600 & {\rev 1120} &   5.8 &   5.8 \\
15  & 26nov04 & {\rev 3336.35965} & 4$\times$600 & {\rev  770} &  -7.5 &   7.8 \\
16  & 27nov04 & {\rev 3337.41602} & 4$\times$600 & {\rev  680} &  -1.6 &   8.8 \\
17  & 30nov04 & {\rev 3340.42430} & 4$\times$600 & {\rev 1000} &  -3.7 &   6.2 \\
18  & 02jul05 & {\rev 3554.64367} & 4$\times$300 & {\rev 1160} &  -8.6 &   5.5 \\
19  & 03jul05 & {\rev 3555.64737} & 4$\times$300 & {\rev  750} &  -6.8 &   7.9 \\
20  & 10jul05 & {\rev 3562.63569} & 4$\times$300 & {\rev  680} &  -0.9 &   8.3 \\
21  & 15jul05 & {\rev 3567.61350} & 4$\times$300 & {\rev  910} &   6.8 &   6.7 \\
22  & 15jul05 & {\rev 3567.63220} & 4$\times$400 & {\rev 1080} &   3.3 &   5.9 \\
23  & 16jul05 & {\rev 3568.56584} & 4$\times$300 & {\rev  490} &  -2.5 &  11.4 \\
24  & 16jul05 & {\rev 3568.63673} & 4$\times$400 & {\rev  910} &  -2.3 &   6.5 \\
25  & 03dec05 & {\rev 3708.44124} & 4$\times$900 & {\rev 1160} &   4.8 &   5.6 \\
26  & 10dec05 & {\rev 3715.24809} & 4$\times$300 & {\rev  890} &  -5.2 &   6.7 \\
27  & 11dec05 & {\rev 3716.41796} & 4$\times$300 & {\rev  860} & -16.5 &   6.9 \\
28  & 12dec05 & {\rev 3717.25230} & 4$\times$300 & {\rev  990} &   9.8 &   6.1 \\
29  & 13dec05 & {\rev 3718.45764} & 4$\times$300 & {\rev  730} & -14.8 &   8.5 \\
30  & 14dec05 & {\rev 3719.27452} & 4$\times$300 & {\rev  830} &  -1.6 &   7.1 \\
31  & 15dec05 & {\rev 3720.43865} & 4$\times$300 & {\rev  620} &   7.4 &   9.2 \\
32  & 18dec05 & {\rev 3723.23341} & 4$\times$300 & {\rev 1100} &   8.2 &   5.8 \\
33  & 19dec05 & {\rev 3724.23705} & 4$\times$300 & {\rev 1200} &   0.7 &   5.5 \\
34  & 20dec05 & {\rev 3725.23694} & 4$\times$300 & {\rev 1010} &  -4.7 &   6.6 \\
\hline	      
\end{tabular} 
\end{center}
\label{logmusicos}
\end{table}

Data were reduced with our own version of the {\sc Esprit} reduction package
\citep{donati1997}. The usual bias and flat-field corrections were applied, as well
as a wavelength calibration with a ThAr lamp. Each echelle order was then
carefully normalised with the continuum package of the {\sc IRAF}
software\footnote{IRAF is distributed by the National Optical Astronomy
Observatory, which is operated by the Association of Universities for Research
in Astronomy (AURA) under cooperative agreement with the National Science
Foundation.}. 

\subsection{Narval observations}

Narval is a fibre-fed echelle spectropolarimeter with a resolving power of
65000, which replaced Musicos at the TBL in 2006. The spectrograph covers
the wavelength domain from 3694 to 10483 \AA.

We collected 23 Stokes V measurements of $\gamma$\,Peg with Narval in
November-December 2007. Their S/N varies between 661 and 1682 at
5000 \AA\ in the I spectrum. See Table~\ref{lognarval}. Nine of the spectra are
saturated in certain wavelength regions. These regions have been rejected. 

Data were reduced with the {\sc Libre-Esprit} reduction package, an
extension of {\sc Esprit} \citep{donati1997} for Narval available
at the telescope, in a manner similar to that used for the Musicos data. 

\subsection{DAO observations}

\begin{table}[!ht]
\caption[]{Journal of 23 Narval observations of $\gamma$\,Peg. The S/N given in
col. 5 is the one measured at 5000 \AA\ in the I spectrum. Col. 6 indicates the
number of lines used in the LSD mask.}
\begin{center}
\begin{tabular}{l@{\ \ }ll@{\ \ }l@{\ \ }r@{\ \ }rrr}
\hline
\hline
\# & Date & mid-HJD    & T$_{\rm exp}$ & S/N & Lines & $B_l$ & $\sigma_{B_l}$ \\  
   & 2007 & $-$2450000 & s             &     &       & G     & G            \\
\hline
1   & 18nov & {\rev 4423.41795} & 4$\times$300 & {\rev 1680} & 744  &  1.0 & 2.4  \\
2   & 18nov & {\rev 4423.43313} & 4$\times$120 & {\rev  990} & 1012 & -0.9 & 1.9  \\
3   & 18nov & {\rev 4423.44134} & 4$\times$120 & {\rev 1060} & 1012 &  3.4 & 1.8  \\
4   & 18nov & {\rev 4423.44951} & 4$\times$120 & {\rev 1070} & 1012 &  2.2 & 1.7  \\
5   & 24nov & {\rev 4429.42376} & 4$\times$120 & {\rev  660} & 1012 &  0.3 & 2.7  \\
6   & 24nov & {\rev 4429.43503} & 4$\times$200 & {\rev  690} & 1012 & -1.3 & 2.6  \\
7   & 28nov & {\rev 4433.39564} & 4$\times$200 & {\rev 1130} & 1012 & -1.4 & 1.6  \\
8   & 04dec & {\rev 4439.31207} & 4$\times$300 & {\rev  790} & 1012 & -0.2 & 2.2  \\
9   & 04dec & {\rev 4439.32835} & 4$\times$300 & {\rev  730} & 1012 & -3.9 & 2.4  \\
10  & 04dec & {\rev 4439.34462} & 4$\times$300 & {\rev  740} & 1012 & -2.2 & 2.4  \\
11  & 12dec & {\rev 4447.36642} & 4$\times$300 & {\rev 1260} & 1012 & -1.4 & 1.4  \\
12  & 12dec & {\rev 4447.38270} & 4$\times$300 & {\rev 1320} & 979  & -0.2 & 1.4  \\
13  & 12dec & {\rev 4447.39898} & 4$\times$300 & {\rev 1340} & 950  & -0.1 & 1.4  \\
14  & 14dec & {\rev 4449.34232} & 4$\times$300 & {\rev 1590} & 791  & -1.1 & 1.9  \\
15  & 14dec & {\rev 4449.35861} & 4$\times$300 & {\rev 1560} & 808  &  0.5 & 1.9  \\
16  & 14dec & {\rev 4449.37489} & 4$\times$300 & {\rev 1480} & 864  &  0.1 & 2.0  \\
17  & 15dec & {\rev 4450.34251} & 4$\times$300 & {\rev 1070} & 1012 &  3.7 & 1.7  \\
18  & 15dec & {\rev 4450.35880} & 4$\times$300 & {\rev 1140} & 1012 & -1.2 & 1.6  \\
19  & 16dec & {\rev 4451.33234} & 4$\times$300 & {\rev  950} & 1012 & -0.1 & 1.9  \\
20  & 16dec & {\rev 4451.34861} & 4$\times$300 & {\rev 1000} & 1012 &  2.8 & 1.8  \\
21  & 18dec & {\rev 4453.33791} & 4$\times$300 & {\rev 1640} & 794  & -1.3 & 2.2  \\
22  & 18dec & {\rev 4453.35963} & 4$\times$300 & {\rev 1660} & 758  & -2.1 & 2.4  \\
23  & 18dec & {\rev 4453.37592} & 4$\times$300 & {\rev 1520} & 848  & -3.0 & 2.3  \\
\hline			
\end{tabular}		
\end{center}		
\label{lognarval}	
\end{table}		
			
\begin{table}[!ht]	
\caption[]{Journal of 18 DAO observations of $\gamma$\,Peg. The reported field values have been measured in the H$\beta$ line. Column 4 provides the number of sub-exposures, individual exposure time as well as total exposure time for each measurement. Column 5 indicates the number of times the plate has been switched per sub-exposure.}
\begin{center}		
\begin{tabular}{l@{\,\,}l@{\,\,}ll@{\,\,}r@{\,\,}l@{\,\,}r@{\,\,}r}
\hline
\hline
\# & Date & mid-HJD	& T$_{\rm exp}$ & S/s. & S/N  & $B_l$  & $\sigma_{B_l}$  \\  
   & 	  & $-$2450000  & s		& & 	 & G	  & G  \\
\hline
1   & 27dec07 & 4461.65079 & 40$\times$30=1200 & 30 & {\rev 1820} & -41 & 38  \\
2   & 27dec07 & 4461.66969 & 80$\times$15=1200 & 30 & {\rev 1820} & -13 & 47  \\
3   & 27dec07 & 4461.68948 & 40$\times$30=1200 &  2 & {\rev 1620} &  62 & 75  \\
4   & 27dec07 & 4461.70234 &  20$\times$30=630 &  6 & {\rev 1100} &  53 & 87  \\
5   & 27dec07 & 4461.71476 &  20$\times$30=600 & 10 & {\rev 1020} &  77 & 87  \\
6   & 7sep08  & 4716.89516 &  30$\times$30=900 & 60 & {\rev 1610} &  84 & 67  \\
7   & 8sep08  & 4717.90738 &  19$\times$18=342 & 60 & {\rev 1240} & -34 & 64  \\
8   & 8sep08  & 4717.91400 &  20$\times$18=360 & 10 & {\rev 1270} &  14 & 79  \\
9   & 8sep08  & 4717.92539 &  20$\times$18=360 & 60 & {\rev 1260} &  -8 & 57  \\
10  & 2dec09  & 5167.70960 & 60$\times$30=1800 & 60 & {\rev 1730} &  10 & 57  \\
11  & 3dec09  & 5168.69853 & 79$\times$30=2370 & 60 & {\rev 2220} &  45 & 38  \\
12  & 4dec09  & 5169.67541 & 100$\times$30=3000& 60 & {\rev 2380} &  28 & 37  \\
13  & 5dec09  & 5170.64841 & 80$\times$60=4800 & 60 & {\rev 2090} &  -6 & 44  \\
14  & 6dec09  & 5171.64258 & 94$\times$30=2820 & 60 & {\rev 1890} & -38 & 53  \\
15  & 26dec09 & 5191.59017 & 100$\times$30=3000& 60 & {\rev 1780} &  60 & 48  \\
16  & 27dec09 & 5192.71002 & 100$\times$30=3000& 60 & {\rev 1370} & 103 & 62  \\
17  & 28dec09 & 5193.66309 & 100$\times$30=3000& 60 & {\rev 1670} & 102 & 50  \\
18  & 3jan10  & 5199.64027 & 99$\times$30=2970 & 60 & {\rev 1550} &  66 & 69  \\
\hline
\end{tabular}
\end{center}
\label{logdao}
\end{table}

\begin{figure}[!ht]
\begin{center}
\resizebox{\hsize}{!}{\includegraphics[clip]{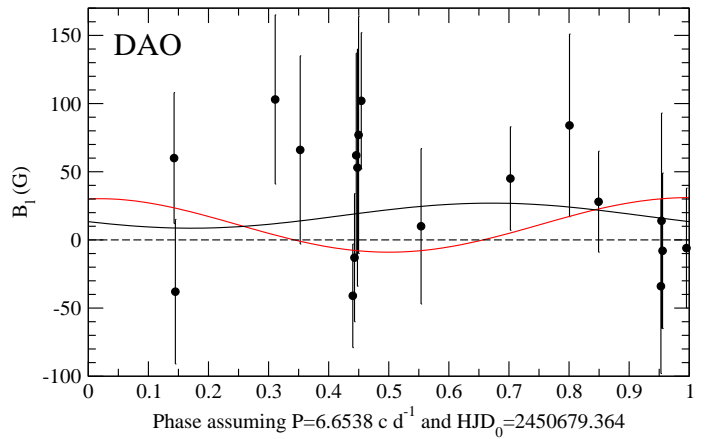}}
\caption[]{Longitudinal field measurements from the H$\beta$ line observed with
DAO. The black solid line shows the best sinusoidal dipole fit to the data, the
dashed line shows a null field, and the red solid line shows the sinusoidal
variation expected from BP07.}
\label{bldao}
\end{center}
\end{figure}

\begin{figure}[!ht]
\begin{center}
\resizebox{\hsize}{!}{\includegraphics[clip]{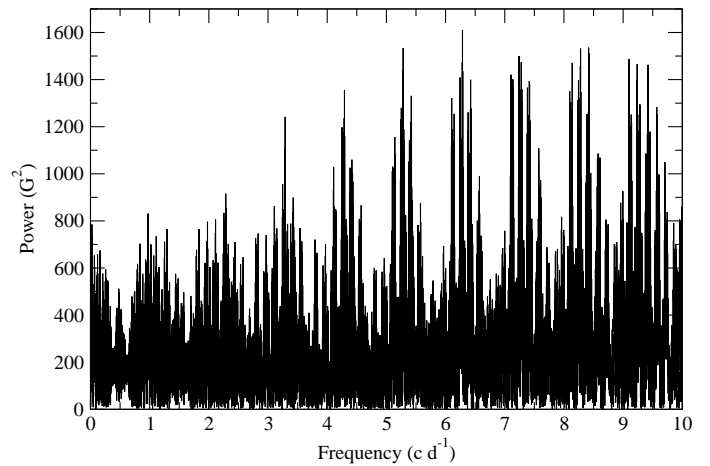}}
\caption[]{Power spectrum of the DAO longitudinal field measurements.}
\label{daofourier}
\end{center}
\end{figure}

The dimaPol spectropolarimeter on the 1.8-m DAO Plaskett telescope was used to
obtain magnetic field measurements in the hydrogen line H$\beta$. The
spectropolarimeter has a resolving power of 10000. Spectra in opposite circular
polarisations, of approximately 250 \AA\ wide and centred on H$\beta$, are
recorded on the CCD. A fast switching liquid crystal wave plate quickly
interchanges the spectra on the detector. Fast switching of the plate combined
with charge shuffling on the CCD significantly reduces instrumental effects and
increases the accuracy of spectropolarimetric measurements. Details about the
instrument and data reduction can be found in \cite{monin2012}.

We collected 18 measurements of $\gamma$\,Peg with dimaPol at the DAO between
2007 and 2010. A single observation of $\gamma$\,Peg consists of between 20 and
100 sub-exposures of 18 to 60 seconds long. Sixty switches per sub-exposure are
typically performed. Some observations obtained in 2007 and 2008 were obtained
with 2 to 30 switches per sub-exposure. The S/N of the measurements
varies between 1000 and 2400 at 5000 \AA\ in the I spectrum.  See
Table~\ref{logdao}.

\section{Spectropolarimetric analysis}\label{analysis}

\subsection{H$\beta$ analysis of DAO data}

\begin{figure*}[!ht]
\begin{center}
\resizebox{0.4\hsize}{!}{\includegraphics[clip]{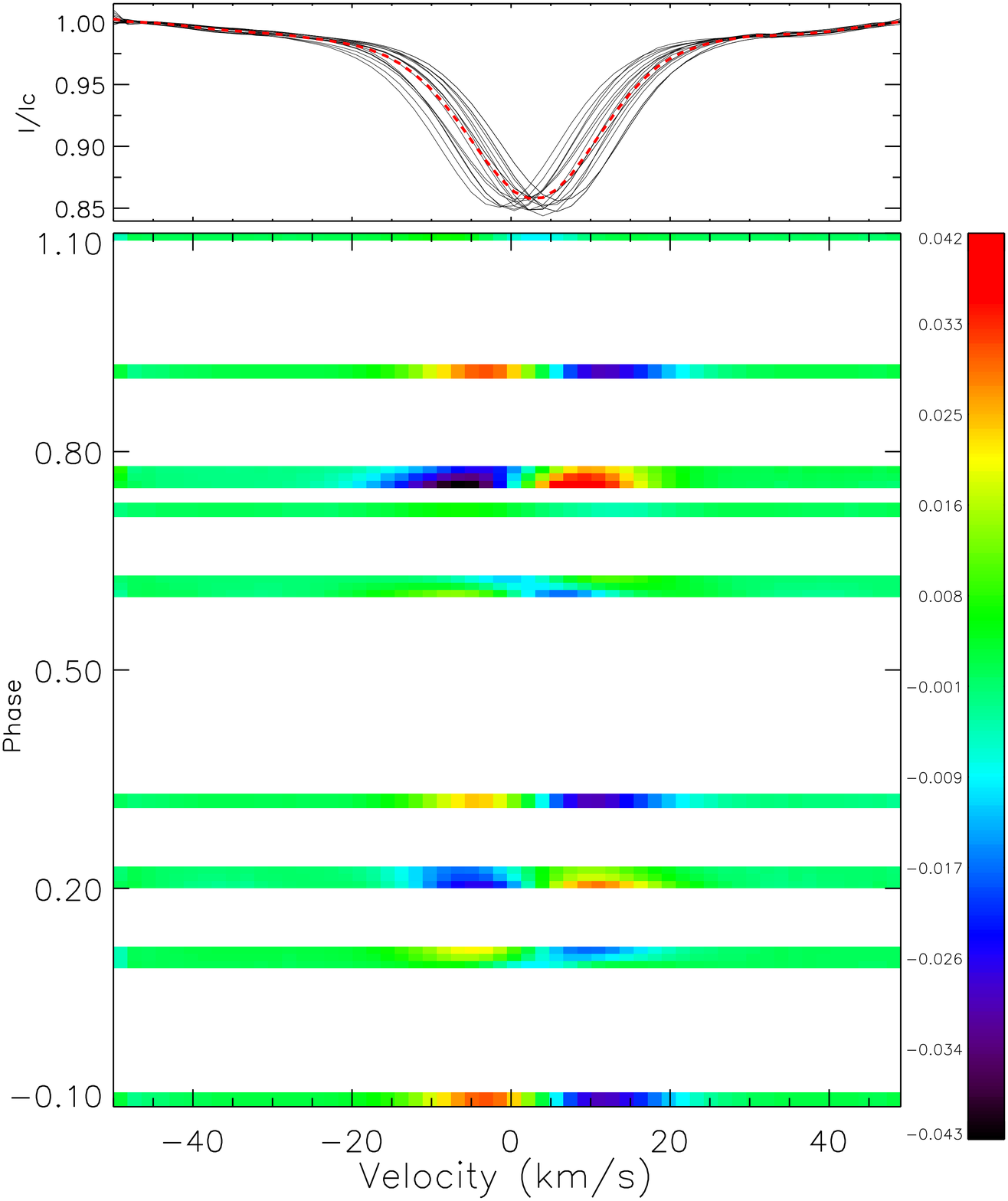}}
\resizebox{0.4\hsize}{!}{\includegraphics[clip]{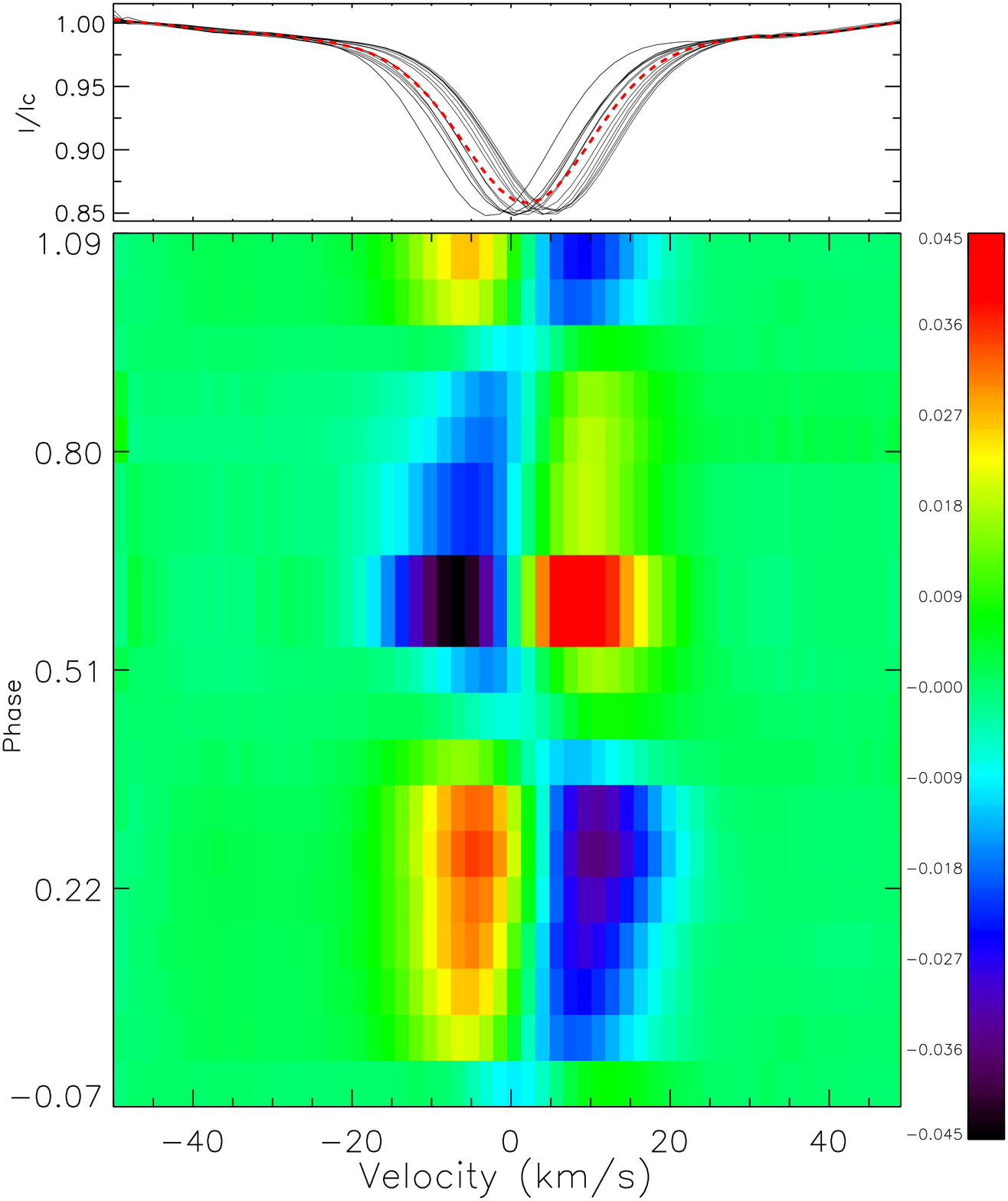}}
\caption[]{Top: LSD I profiles. The mean profile is shown with a dashed red
line. Bottom: dynamical plot of the residuals of the LSD I profiles compared to
the mean LSD I profile, folded in phase with the ephemeris of BP07 (left) and
with the same HJD$_0$ but the main pulsation period P$_{\rm puls}$=0.15175 d
(right). The profiles have been rebinned in bins of 0.01 for the
period of BP07 and 0.06 for the pulsation period.}
\label{dynI}
\end{center}
\end{figure*}

The longitudinal field values $B_l$ extracted from DAO observations have been
obtained by measuring the Zeeman shift between the two opposite circular
polarisations in the core of the H$\beta$ line with the Fourier
cross-correlation technique. This shift is proportional to the
longitudinal field \citep[see e.g.][]{landstreet1992}, with a coefficient of 6.8
kG per pixel for H$\beta$ for dimaPol. The shift is measured using unnormalised
spectra that have not been wavelength calibrated (i.e. in pixel space). We only
apply a chromatic correction so that the continuum shape is similar in both
polarisations. The window of the Fourier cross-correlation has been adjusted to
2.4 \AA\ in order to include the part of the line profile most sensitive to the
magnetic field. See more details about this technique in \cite{monin2012}.

We find that the $B_l$ values are all compatible with 0 within 3$\sigma_{B_l}$.
The field values and their error bars are reported in Table~\ref{logdao}. The
mean longitudinal field obtained from the 18 DAO measurements is $B_l=8\pm14$
G. 

In the top panel of Fig.~\ref{bldao} the longitudinal field values obtained from
the H$\beta$ line of DAO observations are shown folded with P=6.6538 d and
HJD$_0$=2450679.364, the ephemeris suggested by BP07. A sinusoidal dipole fit of
the $B_l$ data is performed and compared to a null field. While we cannot
reproduce the sinusoidal variation expected from BP07 (red line), a dipole fit
with their period but different phasing is possible (black line). The best fit
is a sinusoid centred at 22.5 G with an amplitude $B_0$=13.2 G. It results in a
reduced $\chi^2$=0.86, while the null field (dashed line) has $\chi^2$=0.99.
However, the DAO datapoints appear very scattered around this fit.  Moreover,
considering the distribution of the small number of datapoints over several
years, the periodicity possibly present in these data is very badly constrained.
Therefore, we are also able to find similarly good sinusoidal fits with a
variety of other periods: the Fourier spectrum of the DAO longitudinal field
measurement is shown in Fig.~\ref{daofourier}. As a conseqeunce, the fit of the
DAO data with BP07's period does not appear to be significant.

\subsection{LSD analysis of Musicos and Narval data}

We applied the LSD technique \citep{donati1997} to the Musicos and Narval data.
We first constructed a line mask based on a line template derived from the VALD
database \citep{piskunov1995, kupka1999} with T$_{\rm eff}=22000$ K and $\log
g=3.5$ dex. The template only contains lines with a depth above 1\% of the
continuum level. From the line template we removed all hydrogen lines, lines
that are blended with H lines or interstellar bands, as well as lines that did
not seem to be present in the Narval spectra. We then adjusted the
strength of the remaining lines to fit the Narval observations. This resulted in
masks containing 491 lines for Musicos and between 744 and 1012 lines for
Narval. There are less lines in Musicos spectra because the wavelength range is
smaller. The number of lines for Narval depends on whether we had to discard
parts of the wavelength range where the signal was saturated. This number is
indicated in Table~\ref{lognarval}.

Using these line masks, we extracted LSD Stokes I and V profiles for each
spectropolarimetric measurements. We also extracted null (N) polarisation
profiles to check for spurious signatures, e.g. from instrumental origin or from
stellar pulsations. 

Fig.~\ref{dynI} shows the Narval LSD I profiles folded with the ephemeris
proposed by BP07 and with the main pulsation period (P$_{\rm puls}=0.15175$ d)
found by \cite{handler2009}. We find that the LSD I profiles indeed vary with
the published main pulsation period, but no coherent variation is found with the
period published by BP07.

\begin{figure}[!ht]
\begin{center}
\resizebox{\hsize}{!}{\includegraphics[clip]{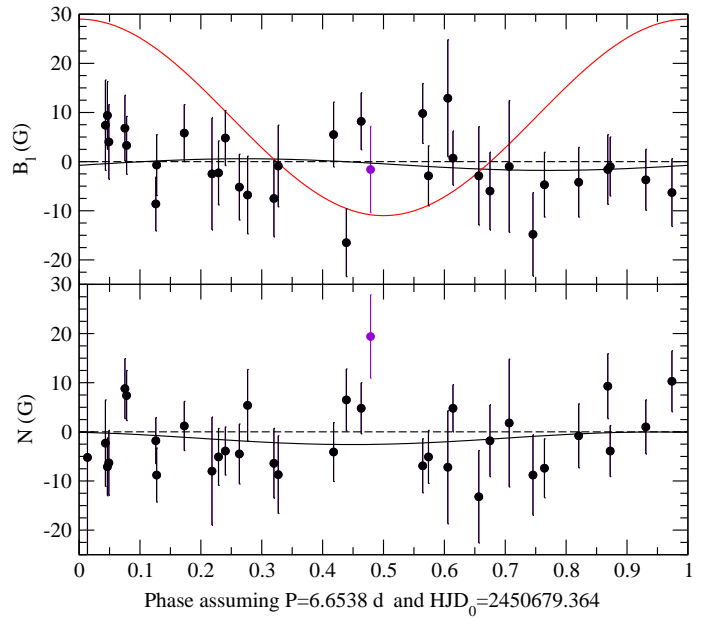}}
\caption[]{Longitudinal field measurements from the LSD Stokes V profiles (top)
and null N measurements (bottom) observed with Musicos. The black solid lines
show the best sinusoidal dipole fit to the data, the dashed lines show a null
field, and the red solid line shows the sinusoidal variation expected from BP07.
The purple point has been discarded for the fit, due to its high N value.}
\label{blmusicos}
\end{center}
\end{figure}

We find that all N profiles are flat, i.e. that the measurements have not been
polluted, and that all Stokes V profiles are also flat, i.e. we do not detect a
magnetic signature in any of the Musicos and Narval measurements. 

We computed longitudinal field ($B_l$) values and their error bars
($\sigma_{B_l}$) from the LSD profiles. The error bars for Musicos and Narval
data are of the order of 8 G and 2 G, respectively. We find that the $B_l$
values are all compatible with 0 within 3$\sigma_{B_l}$. The field values and
their error bars are reported in Tables~\ref{logmusicos} and \ref{lognarval}.
The mean longitudinal field obtained from the 34 Musicos measurements is
$B_l=-0.5\pm1.2$ G. The mean longitudinal field obtained from the 23 Narval
measurements is $B_l=-0.1\pm0.4$ G. These results are not compatible with the
field values proposed by BP07.

\begin{figure}[!ht]
\begin{center}
\resizebox{\hsize}{!}{\includegraphics[clip]{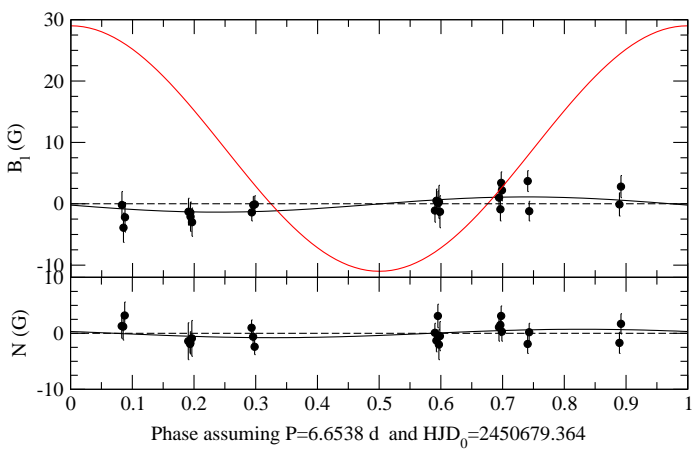}}
\caption[]{Longitudinal field measurements from the LSD Stokes V profiles (top)
and null N measurements (bottom) observed with Narval. The black solid lines
show the best sinusoidal dipole fit to the data, the dashed lines show a null
field, and the red solid line shows the sinusoidal variation expected from
BP07.}
\label{blnarval}
\end{center}
\end{figure}

In the top panel of Figs.~\ref{blmusicos} and \ref{blnarval} the longitudinal
field values obtained from Musicos and Narval observations, respectively, are
shown folded with the ephemeris suggested by BP07. The null N values are also
reported in the bottom panel of each figure. A sinusoidal dipole fit of the
$B_l$ and N data is performed (black solid line) and compared to a null field
(dashed line).

For Musicos data, we rejected one point for the sinusoidal fit, for which the N
value is larger than twice its error bar (see purple point in
Fig.~\ref{blmusicos}). The best sinusoidal fit for $B_l$ has an amplitude of 1.2
G and is centred at $B_0$=-0.6 G. For N the best fit has an amplitude of 1.3 G
and is centred at $B_0$=-1.3 G. We find that the reduced $\chi^2$ for $B_l$
values is $\chi^2$=0.97 for the dipole fit and $\chi^2$=0.93 for the null field.
For the N values it is $\chi^2$=0.96 for the dipole fit and $\chi^2$=0.96 for
the null field. Therefore all four fits are acceptable (i.e. a dipole fit is not
better than no field) and the $B_l$ values do not show a more significant
sinusoidal variation than the N values. 

For Narval data, the best sinusoidal fit for the $B_l$ has an amplitude of 1.24
G and is centred at $B_0$=-0.12 G. For N the best fit has an amplitude of 0.75
G and is centred at $B_0$=0.02 G. We find that the reduced $\chi^2$ for $B_l$
values is $\chi^2$=0.77 for the dipole fit and $\chi^2$=1.01 for the null field.
For the N values it is $\chi^2$=0.87 for the dipole fit and $\chi^2$=0.83 for
the null field. Therefore we find again that all four fits are acceptable  (i.e.
a dipole fit is not better than no field) and the $B_l$ values do not show a
more significant sinusoidal variation than the N values. 

We conclude that we cannot reproduce the sinusoidal variation with an amplitude
of {\rev 20 G centred on $B_0$=11 G} claimed by BP07, shown in red in
Figs.~\ref{blmusicos} and \ref{blnarval}.

\begin{figure}[!ht]
\begin{center}
\resizebox{\hsize}{!}{\includegraphics[clip]{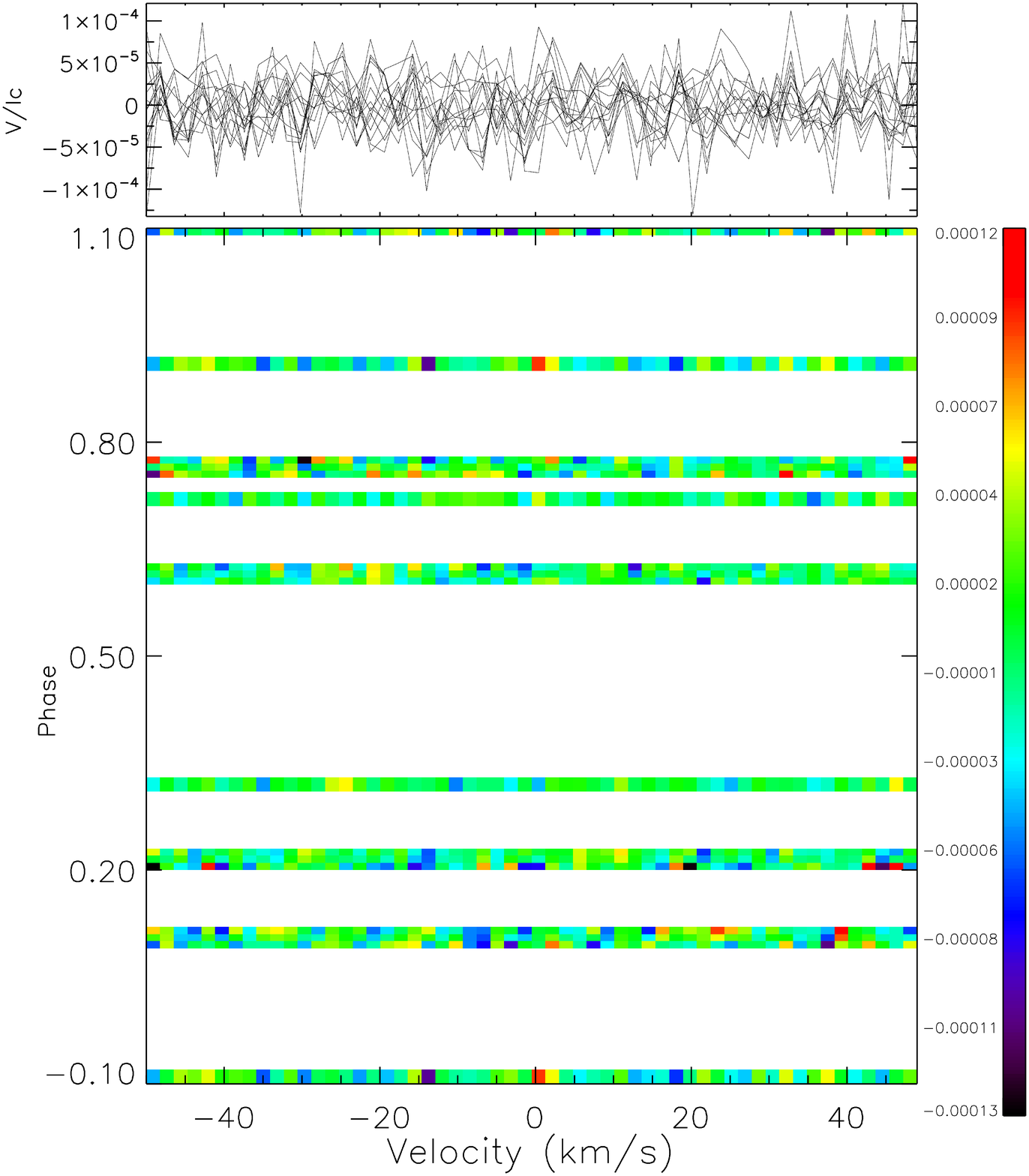}}
\caption[]{Top: LSD V profiles. Bottom: dynamical plot of the LSD V profiles, folded in phase with the ephemeris of BP07. The profiles have been rebinned in bins of 0.01.}
\label{dynV}
\end{center}
\end{figure}

In Fig.~\ref{dynV} we show the LSD V profiles and dynamical plot of these
profiles folded in phase with the ephemeris of BP07 rebinned
in bins of 0.01. We see no signatures in the LSD V profiles and no
travelling features in the dynamical plot.

\subsection{Upper limit on field strength}\label{limit}

To derive an upper limit on the strength of a magnetic field which could have
remained undetected in our data, we used the best data available, i.e. the
Narval dataset.

For various values of the polar magnetic field $B_{\rm  pol}$, we calculated
1000 oblique dipole models of each of the 23 Stokes V profiles with random
inclination angle i and obliquity angle $\beta$, random rotational phase, and a
white Gaussian noise with a null average and a variance corresponding to the S/N
ratio of each Narval profile. To calculate these oblique dipole models, we used
Gaussian local intensity profiles with a width calculated according to the
resolving power of Narval and a thermal broadening. This broadening and the
depth of the intensity profile was determined by fitting the observed LSD I
profiles. We then calculated local Stokes V profiles assuming the weak-field
case and integrated over the visible hemisphere of the star. We obtained
synthetic LSD Stokes V profiles, which we normalised to the intensity continuum.
We used the mean Land\'e factor and wavelength from each LSD profile.

\begin{figure}[!ht]
\begin{center}
\resizebox{\hsize}{!}{\includegraphics[clip]{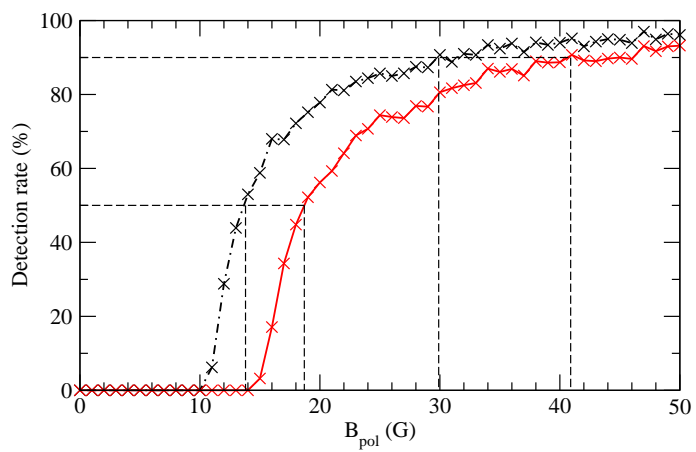}}
\caption[]{Chances that an oblique dipolar magnetic field would have been
detected in $\gamma$\,Peg, as a definite (red solid line) or marginal (black
dashed-dotted line) detection, in at least one of the 23 Narval measurements
according to the strength of the dipolar magnetic field. The thin dashed lines
indicate the 50\% and 90\% detection rate.}
\label{upperlimit}
\end{center}
\end{figure}

We then computed the probability of detection of a field in this set of models
by applying the Neyman-Pearson likelihood ratio test \citep[see
e.g.][]{helstrom1995,kay1998,levy2008} to decide between two hypotheses, $H_0$
and $H_1$, where $H_0$ corresponds to noise only, and $H_1$ to a noisy
simulated Stokes V signal. This rule selects the hypothesis that maximises the
probability of detection while ensuring that the probability of false alarm
$P_{\rm FA}$ is not larger than a prescribed value considered acceptable.
Following values usually assumed in the literature on magnetic field detections
\citep[e.g.][]{donati1997}, we used  $P_{\rm FA} = 10^{-5}$ for a definite
magnetic detection and $P_{\rm FA} = 10^{-3}$ for a marginal magnetic detection.
We then calculated the rate of detections among the 1000 models for each of the
23 profiles depending on the field strength. The definite and marginal
detections in at least one of the 23 LSD profiles rate curves are plotted in
Fig.~\ref{upperlimit}. 

This translates into an upper limit for the possible non-detected polar field
strength of $B_{\rm  pol}=30$ and 41 G for a 90\% chance of marginal or definite
detection, respectively, in at least one of the 23 LSD profiles. This value is
$B_{\rm pol}=14$ and 19 G for a 50\% chance of marginal or definite detection,
respectively. In other words, an oblique dipolar magnetic field above $\sim$40 G
at the poles should have been detected.

\section{Discussion}\label{discuss}

\subsection{Dipolar magnetic field}

The data obtained with Narval leave no doubt that $\gamma$\,Peg does not host a
$B_{\rm pol}=570$ G dipolar field as suggested by BP07. This confirms the
non-detection published by previous authors based on single measurements
\citep{schnerr2008, silvester2009} and obtained from our older series of Musicos
data presented here. The mean longitudinal field value obtained from metallic
lines with Narval is $B_l=-0.1\pm0.4$ G and the upper limit on the polar
strength of a non-detected field is $\sim$40 G for a 90\% chance of definite
detection. 

In addition, although the DAO data provide longitudinal field values with higher
error bars than LSD profiles extracted from Musicos and Narval data, the lower
resolution of dimaPol and its fast switching device allow to check for the
presence of a magnetic field in a single broader hydrogen line. The DAO
measurements of magnetic field in the H$\beta$ line confirms the non-detection
result. 

Therefore we conclude that $\gamma$\,Peg does not host the kind of stable
magnetic fossil field observed in 7\% of massive and intermediate-mass stars
\citep{wade2013}.

\subsection{Variability}

A period around 6.6 d appears in the magnetic measurements collected by BP07
{\rev and cannot be ruled out} in our DAO {\rev data. Its} 1-d alias
(f$\sim$0.85 c d$^{-1}$) lies within the range of g-mode frequencies observed in
$\gamma$\,Peg. Variability due to pulsations can easily appear in magnetic
measurements if the exposure time is longer than about 1/20 of the pulsation
period. In the case of the observed g-modes, this would correspond to exposures
longer than $\sim$5000 seconds. BP07 do not indicate the duration of their
exposures, therefore it is not possible to check whether the length of their
exposures is the reason for the detection of this period in their magnetic
measurements. 

The much shorter and higher S/N Narval exposures do not show the variability
around 6.6 d. Some of our DAO exposures are of the order of $\sim$5000-second
duration. However, the rapid switching of the wave plate in dimaPol should avoid
a pollution of the measurements by stellar variations such as pulsations.  

In addition, while we are able to find a sinusoidal fit to the DAO data with
BP07's period but a completely different phasing (see Fig.~\ref{bldao}), we are
also able to find similarly good sinusoidal fits with a variety of other periods
(see Fig.~\ref{daofourier}). Therefore the fit to DAO data with BP07's period
does not appear to be significant.

\subsection{Vega-like magnetic field}

Very weak magnetic fields have been detected in a few A stars
\citep[e.g.][]{petit2011}. The first such field was discovered in the star Vega
\citep{lignieres2009,petit2010} and this is why they are called "Vega-like
fields". The existence of these very weak fields in A stars is still debated and
no such field has ever been detected in an O or B star.

\begin{figure}[!ht]
\begin{center}
\resizebox{\hsize}{!}{\includegraphics[clip]{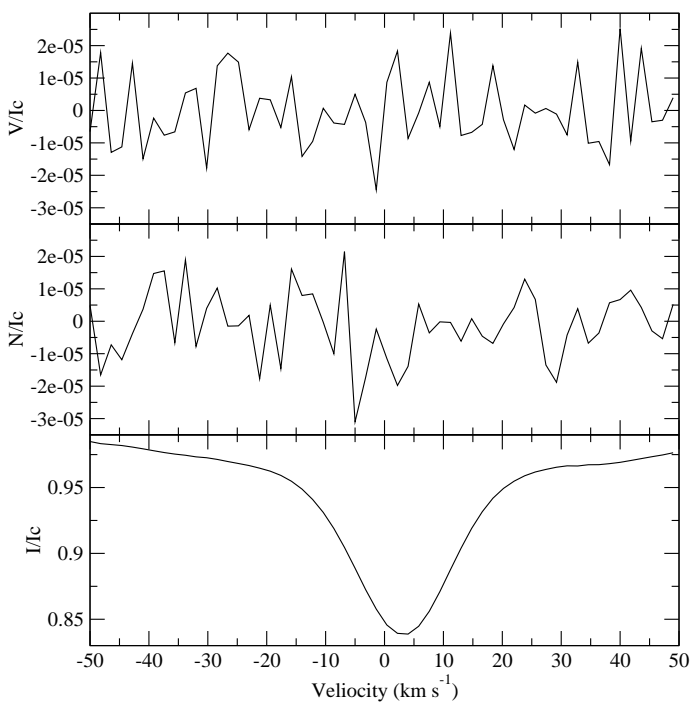}}
\caption[]{Average Narval LSD Stokes V (top), null polarisation (middle) and
intensity (bottom) profiles.}
\label{averageLSD}
\end{center}
\end{figure}

For Vega, the measured longitudinal field is $-0.6\pm0.3$ G
\citep{lignieres2009}. For Sirius, the measured Vega-like longitudinal field is
$0.2\pm0.1$ G. These measurements, although weak (a few tens of a Gauss), are
not compatible with 0. Moreover, the weak signature of these fields is observed
in the average LSD Stokes V profiles of these stars \citep[see Fig.~1
in][]{petit2011}.

For $\gamma$\,Peg, with our best measurements (i.e. Narval data) and using the
same method as \cite{lignieres2009} and \cite{petit2011}, we find a longitudinal
field of $-0.1\pm0.4$ G. Contrary to Vega and Sirius, this field is compatible
with 0. Moreover, we find no signature in the average LSD Stokes V profile, as
shown in  Fig.~\ref{averageLSD}. Therefore we can also exclude that the
longitudinal field has been averaged out over the stellar surface.

We thus conclude that $\gamma$\,Peg does not host a Vega-like field.

\subsection{On the existence of Vega-like fields in most OBA stars}

In magnetic massive stars with strong (above a few hundreds Gauss) magnetic
fields, the field is thought to be of fossil origin. The primordial field
evolves to an equilibrium state: a twisted torus inside the star, that appears
most of the time as a simple oblique dipole field at the surface
\citep{braithwaite2006,duez2010}.

After the discovery of very weak magnetic fields in Vega and Sirius, two
scenarios have been proposed to explain the existence of such weak fields.
First, \cite{auriere2007} proposed that in the case of Vega-like stars, the
magnetic field has reached the equilibrium but the field amplitude was too weak
to freeze a possible differential rotation. As a consequence, this differential
rotation created a strong toroidal field, which became unstable because of the
Tayler instability \citep{tayler1973}. This destroyed the stable field
configuration. Second, \cite{braithwaite2013} proposed that Vega-like stars are
following the same process as the strongly magnetic ones towards a magnetic
equilibrium, but, because their field is much weaker, the time necessary to
reach this equilibrium is longer than the current age of the star and the field
is currently still evolving. If the time to reach the equilibrium is longer that
the lifetime of the star, it is called a failed fossil field. Both scenarios  
conclude that all massive and intermediate-mass stars should host a magnetic
field: either a strong stable fossil field (as observed in 7\% of the stars
according to the MiMeS survey, see \citealt{wade2013}) or a weak Vega-like field
for all the other stars.

However, we find that $\gamma$\,Peg hosts neither a strong stable magnetic field
nor a weak field similar to Vega or Sirius. This raises questions about the
above scenarios. In particular, in order to work, both scenarios require an
initial seed field, coming from the molecular cloud from which the star was
formed. If the star did not capture a field during its collapse, no dynamo can
develop during the pre-main sequence convective phase and there will be no
relaxation towards an equilibrium fossil field in the radiative envelope later
on. One can then wonder why the few A stars observed with deep
spectropolarimetry so far did show a Vega-like field and thus had a seed field,
and the B star $\gamma$\,Peg does not. If $\gamma$\,Peg did capture a seed
field, it seems to have been destroyed, which cannot be explained by the above
two scenarios. 

To test the Vega-like field scenarios further, it would be interesting to study
more bright B stars with very deep spectropolarimetry. Moreover, the A stars for
which a Vega-like field was discovered so far are rather peculiar objects: Vega
is a very rapid rotator seen pole-on and a pulsator, and Sirius is an Am star
and a binary. It would be interesting to check for the presence of Vega-like
fields in ``normal'' A stars. 

\subsection{Implication for stellar evolution models}

The non-detection of a magnetic field in the early B star $\gamma$\,Peg with a
very low detection threshold implies that current stellar evolution models
without magnetic fields might be a good approximation of most massive stars
\citep{meynet2000,maeder2000}. A very weak field, such as a Vega-like field,
could make a great difference in the evolution of a massive star and it was thus
important to check whether these fields are indeed present in all massive and
intermediate-mass stars. Although our work is only based on one star, we have
shown that the predictions by the Vega-like fields theories are too crude and
there is no need to include these weak fields in all evolution models of massive
stars. 

\section{Conclusions}

We have performed extremely sensitive magnetic measurements of the early B star
$\gamma$\,Peg. We found that it does not host the several hundreds Gauss field
claimed by BP07. It also does not host a very weak Vega-like
magnetic field. 

Our results show that, while Vega-like fields may exist in A stars, their
existence in hotter stars and in most OBA stars can be questioned. Very deep
magnetic observations of other bright OBA stars would allow to test further and
constrain the Vega-like theories for massive stars.

\begin{acknowledgements}
{\rev We thank the Musicos observers H. Henrichs, V. Geers and N. Boudin, the
Narval observers J. Guti\'errez-Soto and F. Cochard, as well as the TBL service
observation team.} CN wishes to thank the Programme National de Physique
Stellaire (PNPS) for their support and acknowledges support from the ANR (Agence
Nationale de la Recherche) project Imagine. CN thanks G. Meynet for fruitful
discussions, which led to the Narval observations presented here. This research
has made use of the SIMBAD database operated at CDS, Strasbourg (France), and of
NASA's Astrophysics Data System (ADS).
\end{acknowledgements}

\bibliographystyle{aa}
\bibliography{gampeg_Neiner_revised}

\end{document}